# Multi-GeV electron bunches from an all-optical laser wakefield accelerator

B. Miao[1,*], J.E. Shrock[1,*], L. Feder[1,*], R. C. Hollinger[2], J. Morrison[2], R. Nedbailo[2], A. Picksley[3], H. Song[2], S. Wang[2], J.J. Rocca[2,4], and H.M. Milchberg[1,5,†]

[1]*Institute for Research in Electronics and Applied Physics & Dept. of Physics, University of Maryland, College Park, MD 20742, USA*
[2]*Department of Electrical and Computer Engineering, Colorado State University, Fort Collins, CO 80523, USA*
[3]*John Adams Institute for Accelerator Science and Department of Physics, University of Oxford, OX1 3RH, UK*
[4]*Physics Department, Colorado State University, Fort Collins, CO 80523, USA*
[5]*Department of Electrical and Computer Engineering, University of Maryland, College Park, MD 20742, USA*

We present the first demonstration of multi-GeV laser wakefield acceleration in a fully optically formed plasma waveguide, with an acceleration gradient as high as 25 GeV/m. The guide was formed via self-waveguiding of <15 J, 45 fs (< ~300 TW) pulses over 20 cm in a low density hydrogen gas jet, with accelerated electron bunches simultaneously driven up to 5 GeV in a milliradian divergence quasi-monoenergetic peak of relative energy width ~15% and charge of at least ~10 picocoulombs. Energy gain is inversely correlated with on-axis waveguide density in the range $N_{e0} = (1.3 - 3.2) \times 10^{17}$ cm$^{-3}$. We find that shot-to-shot stability of bunch spectra and charge are strongly dependent on the pointing of the injected laser pulse and gas jet uniformity. We also observe evidence of pump depletion-induced dephasing, a consequence of the long optical guiding distance.

## I. INTRODUCTION

Among the compact techniques for laser-driven electron acceleration [1-3], laser-wakefield acceleration (LWFA) in plasmas has achieved the highest energy gains by far [4-6]. For application to light sources and to high energy physics, a key goal has been the development of a high repetition rate ~10 GeV-scale laser-driven accelerator module. For a ~ 1 TeV-scale centre of mass lepton collider, the sequential staging [7] of dozens of these modules is envisioned [8].

Achieving multi-GeV electron bunches with LWFA requires maintaining the laser intensity at a level sufficient to drive a relativistic plasma wake over distances corresponding to many Rayleigh ranges of the focused pulse. This demands some type of optical guiding, either relativistic self-guiding [9,10] or preformed plasma waveguides that are laser-generated [11] or formed by a capillary discharge [12, 13]. For multi-GeV acceleration in a single stage, low density plasmas are needed: the electron energy gain, accounting for dephasing and depletion, scales with electron density ($N_e$) and laser intensity ($I [W/cm^2] = 1.4 \times 10^{18} a_0^2 (\lambda[\mu m])^{-2}$) as $\Delta W/mc^2 \sim a_0^r N_{cr}/N_e$ [14], where $a_0$ is the normalized vector potential and $N_{cr}$ is the critical density, and where $r = 2$ in the quasilinear regime ($a_0 > \sim 1$) and $r = 1$ in the 3D nonlinear blowout regime ($a_0 \gg 1$). At the low plasma densities $N_e \sim 10^{17}$ cm$^{-3}$ consistent with multi-GeV acceleration, relativistic self-guiding (in the blow-out regime) requires at least petawatt laser powers [5, 6]. The required laser power per GeV of acceleration is significantly reduced for preformed plasma waveguides, where one can operate in the quasilinear regime [4, 14, 15]. The $\Delta W$ scaling predicts that a dephasing-limited ~10 GeV acceleration stage driven by a guided laser pulse with $a_0 > \sim 1$ requires $N_e \sim 10^{17}$ cm$^{-3}$. This corresponds to dephasing lengths $L_d < \sim 1$ m, which sets the approximate length $L_{guide}$ required for the plasma waveguide. Plasma waveguides also enable independent control of guided laser mode structure and propagation characteristics, as well as control over dephasing, depletion, and phase matching in electron acceleration [16-20].

---

[*] These authors contributed equally to this work.
[†] milch@umd.edu

In this paper we present the first demonstration of an all-optical multi-GeV laser wakefield accelerator where laser pulses both generate the plasma waveguide and drive the wakefield acceleration. We observe quasi-monoenergetic electron bunches of energy up to 5 GeV with milliradian beam divergence and ~15% energy spread, with charge of at least ~10 pC in the highest energy peaks. The accelerator wake buckets are injected by electrons from tunneling ionization of He-like nitrogen (N$^{5+}$) [21,22]. Energy gain is found to be inversely correlated with on-axis waveguide density in the range $N_{e0} = (1.3 - 3.2) \times 10^{17}$ cm$^{-3}$. We find that shot-to-shot stability of bunch spectra and charge are strongly dependent on the pointing of the injected laser pulse and gas jet uniformity. We also observe evidence of pump depletion-induced dephasing, a consequence of the long optical guiding distance.

The first laser-generated plasma waveguides [11] relied on cylindrical shock expansion of a Bessel beam-heated plasma to form both the core and cladding of the waveguide. To generate sufficiently large plasma pressure gradients to drive this process required inverse Bremsstrahlung (IB) heating of plasma densities $> \sim 10^{19}$ cm$^{-3}$ to temperatures of tens of eV, typically using ~100 ps pulses requiring ~100 mJ per cm of waveguide generated [23]. However, the plasma density required for a ~10 GeV acceleration stage is 2 orders of magnitude lower and unsuitable for IB heating. As an alternative, the use of optical field ionization (OFI) by ultrashort laser pulses has recently been explored [24-29], motivated by advances leading to cheaper and higher energy short pulse capability. The advantage of OFI is that ionization of the working gas is purely dependent on local laser intensity and *not* gas density. The disadvantage is that electron heating by OFI is limited to electron temperatures comparable to the electron ponderomotive energy in the laser field at the ionization threshold of the gas, less than ~10 eV in hydrogen. For long, low density OFI plasmas, such as those heated with a Bessel beam, the resulting pressure gradient is too weak to drive a shock wall-based plasma cladding sufficient to confine an optical pulse without dominant leakage losses ([28-30]).

Recently, we have experimentally demonstrated two techniques for optical generation of meter-scale low density plasma waveguides, the "2-Bessel" method [30] and "self-waveguiding" [31], with the latter previously demonstrated [32] on few millimeter long channels. While both methods generate low density guides with negligible leakage losses by imprinting the waveguide cladding via OFI, they differ in the details of how the cladding is generated. Both methods use an initial J$_0$ Bessel beam pulse to generate a fully ionized $< \sim 10$ eV hydrogen plasma on axis. This plasma expands radially, snowplowing the peripheral neutral gas into a cylindrical shell of enhanced density, with the central density dropping by up to $\sim 10 \times$. We call this J$_0$-prepared profile the "refractive index structure". In the 2-Bessel method, the intense ring of a few-nanosecond-delayed high order Bessel beam pulse (J$_8$ and J$_{16}$ in [30]) ionizes the peripheral neutrals, forming the plasma cladding. In self-waveguiding, the leading edge of a high intensity pulse injected into the index structure ionizes the neutral shell, forming the cladding on the fly. Self-waveguiding is initially easier to implement than the 2-Bessel method, and we have employed it for the experiments of this paper. But for future experiments requiring improved energy efficiency and additional control of waveguide parameters, the 2-Bessel method remains attractive. The two methods are compared in ref. [31]. We note that the results of earlier guiding experiments [28, 29] on low density OFI-heated plasma channels have been reinterpreted [33] in terms of self-waveguiding, where evidence that a guided pulse could generate additional ionization in its wings had been earlier considered in refs. [34, 35].

## II. EXPERIMENT

The experimental setup is shown in Fig. 1. The laser used in the experiments is the ALEPH laser at Colorado State University [36]. A LWFA drive pulse P1 ($\lambda = 800$ nm, $\tau = 45$ fs FWHM, energy <15 J) was focused by an $f/25$ off-axis paraboloid into the refractive index structure generated in a 20 cm long gas jet by a 0.5 J, 75 fs, zero order Bessel beam ($J_0$) pulse, P2, which was phase corrected by a deformable mirror [30, 31] and then compressed by a separate pulse compressor.

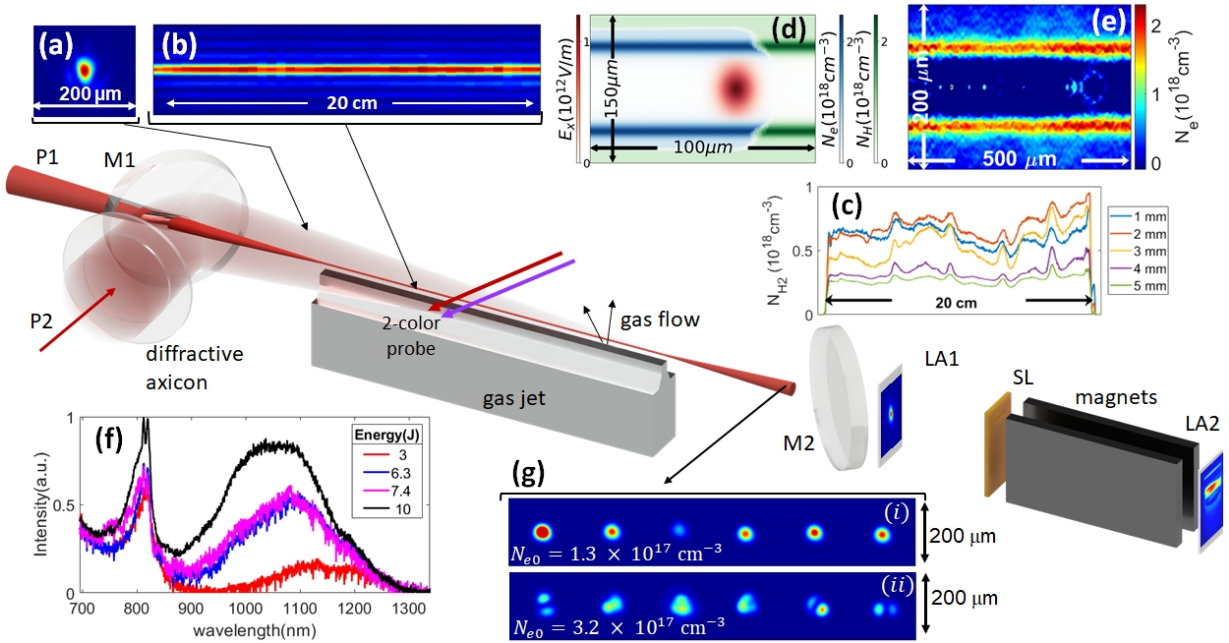

**Figure 1.** Experimental setup. *LWFA drive laser pulse* (P1): ($\lambda = 800$ nm, $\tau_{\text{FWHM}}$ =45 fs, energy <15 J), focused by an $f/25$ off-axis paraboloid through a 9.5 mm diameter, 45° hole in mirror M1. *Index-structuring pulse* (P2): ($\lambda = 800$ nm, $\tau_{\text{FWHM}}$ =75 fs, energy 0.5 J) $J_0$ Bessel beam pulse formed by 4-level transmissive/diffractive axicon, forming a 20 cm long plasma by OFI in the working gas 2.5 ns in advance of P1. The plasma expands radially, forming an elongated refractive index structure: a low density plasma on axis surrounded by an enhanced density annular shell of neutral gas. *Two colour interferometer probes*: for measuring $N_e$ and neutral gas density profiles (see Appendix A). M2: pickoff mirror for guided mode imaging. LA1: Lanex fluorescing screen for full electron beam profile imaging. *Magnetic spectrometer*: 1 mm entrance slit (SL), 30-cm long permanent magnet array (field 0.85 T), Lanex fluorescing screen for electron energy spectrum (LA2) (see Appendix A). *Gas jet*: Mach 4 supersonic nozzle, orifice length 20 cm, fed by 5 solenoid valves backed by pure $H_2$ or a 95/5% $H_2/N_2$ mixture at backing pressure 200-500 psi. Inset panels: **(a)** Focal profile of P1. **(b)** Longitudinal scan of the $J_0$ Bessel beam (P2) focus. **(c)** Axial profiles of gas density vs. height above the nozzle (Appendix A). **(d)** Simulation using the particle-in-cell code FBPIC ([37], Appendix C) of self-waveguiding in hydrogen refractive index structure. **(e)** Plasma waveguide profile interferometrically measured ~1 ps after passage of self-waveguided pulse. **(f)** Guided laser spectra at waveguide exit vs. input pulse energy. **(g)** Effect of shot-to-shot P1 pointing fluctuations on guided mode for (*i*) low density guide and (*ii*) higher density guide.

The drive pulse P1 is focused through a hole in mirror M1, with its beam waist located at the entrance of the index structure; a vacuum mode image is shown in Fig. 1(a). P2 is generated by passing a 5.5 cm diameter super-Gaussian pulse through a 4-level transmissive/diffractive axicon (fused silica substrate, 0.5 mm thick) which converts the 0.9 J input to a $J_0$ beam of energy 0.5 J. The beams forming pulses P1 and P2 are split upstream in the laser chain, prior to their respective compressors, using several fixed ratio beam splitters. An axial ($z$) imaging scan of the $J_0$ beam intensity profile is shown in Fig. 1(b). The diffractive axicon simplifies the experimental geometry and enables co-propagation of P1 and P2, in contrast to counter-propagation necessitated by our prior use of reflective axicons [30, 31]. The rays of the $J_0$ pulse approach the optical axis at angle $\gamma = 2.3°$; the 65cm long focus longitudinally overfills and fully ionizes (via OFI) a 20 cm long column in the gas sheet 3 mm above a Mach 4 supersonic nozzle fed by 5 high pressure pulsed solenoid valves fed with pure $H_2$ or a 95/5% $H_2/N_2$ gas mixture. Axial profiles of $H_2$ density are shown in Fig. 1(c) for various heights above the nozzle, measured as described in Appendix A. The gas density at the ends of the jet sharply transitions to vacuum over ~3 mm. The bumps in the density profiles are due to slight variations in the nozzle orifice width along $z$ and a local intensity bump in the $J_0$ beam focal line. Electron spectra are measured by a $0.75 - 6.5$ GeV range magnetic spectrometer consisting of a 30-cm long permanent magnet array (field 0.85 T)) with a 1 mm entrance slit 3 m from the plasma waveguide exit (see Appendix A).

Figure 1(d) shows a particle-in-cell simulation using the code FBPIC ([37], Appendix C) of the self-waveguiding process: the leading edge of the pulse injected into the index structure forms the plasma waveguide cladding as it propagates (left to right) into the index structure. An interferometric measurement (Appendix A) of the plasma density profile ~ 1 ps after self-waveguiding is shown in Fig. 1(e), where the enhanced plasma density cylinder is the cladding generated by the self-waveguided pulse. The P1 injection delay of 2.5 ns after P2 is chosen so that the $1/e^2$ intensity radius of the lowest order mode of the formed plasma channel matches P1's $1/e^2$ intensity spot radius. The P1 energy leakage from the index structure before self-waveguiding is established is small: for a ~10 J, ~50 fs pulse, the hydrogen ionization threshold of $10^{14}$ W/cm$^2$ is reached at $r = w_{ch} = 30$ μm, ~100 fs before the peak of the pulse. For the guides generated by self-waveguiding in these experiments, we estimate a total cost (in the $J_0$ beam and the self-waveguiding beam) of 15-20 mJ/cm, based on scaling from [31] and simulations (see later discussion). While the self-waveguiding energy cost is small for a 10 J-scale pulse, the laser energy invested in plasma waves can be substantial. The experimental signature of this is increasing energy in the red shifted wings [3] of guided pulses at increasing energy, as seen in Fig. 1(f). The peak of the red wing trending bluer with higher laser pulse energy may be due to pulse lengthening accompanying depletion (see Appendix C).

Waveguide throughput (laser energy exiting the waveguide divided by energy in the P1 focal spot) was measured by integrating CCD camera images of the P1 and guided modes, with the camera energy response calibrated by imaging the P1 focus with known laser energy and calibrated neutral density filters, and adjusted for the pixel spectral response. In this experiment, throughput of only high energy P1 shots could be measured because of several fixed P2/P1 energy ratios and the need for ~2 J of pre-compressed laser energy for the index-structuring (P2) beam. Under these conditions, we measured guided pulse throughputs of <40% depending on laser energy, waveguide density, and P1 pointing, with plasma wave excitation responsible for most of the reduced transmission. This is supported by simulations discussed later.

For fixed nominal laser and waveguide parameters, a major source of shot-to-shot variation in accelerated electron bunches is fluctuating alignment of the drive pulse P1 into the refractive index structure generated by P2. The index structure's transverse position is relatively stable from shot to shot (centroid standard deviations $\sigma_x \sim \sigma_y \sim 2$ μm), as it is mainly determined by transverse positioning of the diffractive axicon and not by variations in P2 pointing. P1 pointing fluctuations, however, are $\sigma_x \sim \sigma_y \sim 9$ μm owing to a longer effective lever arm. The effect of these fluctuations on the waveguide exit mode is shown in Fig. 1(g) for multiple 8 J shots in waveguides with (*i*) $N_{e0} = 1.3 \times 10^{17}$ cm$^{-3}$ and (*ii*) $N_{e0} = 3.2 \times 10^{17}$ cm$^{-3}$. At the lower density, the waveguide supports only the fundamental mode $(p, m) = (0,0)$ (where $p$ and $m$ are the radial and azimuthal mode indices), with P1 pointing fluctuations affecting only the mode peak intensity. At the higher density, the waveguide can also support the $(0,1)$ mode, which accounts for shot-to-shot asymmetry in the waveguide exit beam. Here the $1/e$ power leakage distance of the $(0,0)$ mode is $L_{1/e}^{(0,0)} > 2$ m, while $L_{1/e}^{(0,1)} \sim 1$ m (see Appendix B), showing that off-axis coupling will lead to asymmetric mode contributions that do not leak out of the guide.

## III. RESULTS AND DISCUSSION

The effect of P1 pointing variation on the accelerated electron bunch beam profile and energy spectrum are shown in Fig. 2 for 43 consecutive shots with pulse energy 15 J ± 10%. Here, $N_{e0} = 3.2 \times 10^{17}$ cm$^{-3}$ (as in Fig. 1(g)(ii)) for a dephasing length $L_d = \lambda_p/2(1 - \beta_\phi) \sim 10$ cm, shorter than $L_{guide} = 20$ cm, where $\lambda_p \approx 60$ μm is the plasma wavelength, $\beta_\phi c = v_g - v_{etch}$ is the effective plasma wave velocity, $v_g$ is the laser group velocity and $v_{etch} = (N_{e0}/N_{cr})c$ is the velocity at which the pulse's leading edge erodes backward from pump depletion into the plasma wave [39]. While most injected pulses are guided (row (a)), the presence of a transmitted mode does not guarantee generation of a high quality electron beam. More important is the quality of

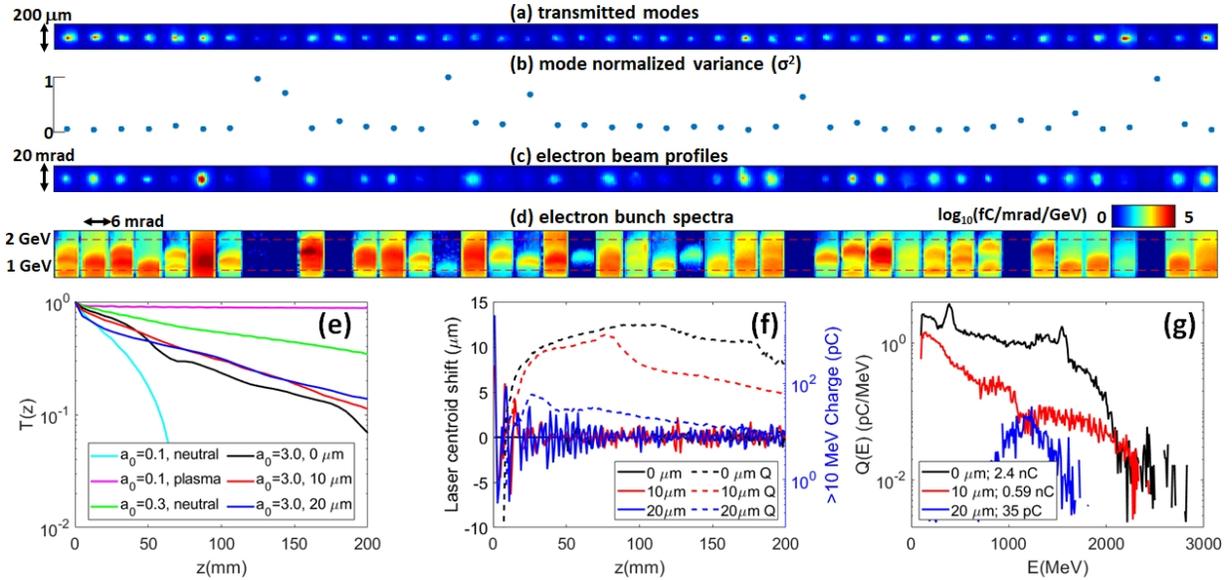

Figure 2. (a) Plasma waveguide exit modes for 43 consecutive shots. Laser 15 J, $\tau = 45$ fs. (b) Normalized mode second moment $\sigma^2$ for each shot. (c) Associated electron beam profiles measured at Lanex screen LA1 in Fig. 1, and (d) associated angle resolved electron bunch spectra. Particle-in-cell simulations using WarpX ([38], Appendix C): (e) Guided pulse energy transmission for three values of $a_0$ and P1 coupling offset. (f) Guided mode centroid oscillation and accelerated charge > 10 MeV for the three P1 coupling offsets ($a_0 = 3.0$). (g) Accelerated bunch spectra for the three P1 coupling offsets ($a_0 = 3.0$).

guiding, which we assess by the second moment of the mode intensity profile $\sigma^2 = (\int dA\, I(\mathbf{r}))^{-1} \int dA |\mathbf{r} - \mathbf{r}_c|^2 I(\mathbf{r})$ of the transmitted intensity profile, where $\mathbf{r}_c$ is the mode

centroid. These are normalized and plotted in row (b), showing a clear correlation to the electron bunch quality variation in beam profile (row (c)) and energy (row (d)). It is important to emphasize that for all experiments of this paper, there was no observed electron acceleration for waveguides generated in pure $H_2$ gas; only the $H_2/N_2$ gas mix yielded LWFA bunches, showing that our accelerator is purely ionization-injected.

To provide physical insight on the effects of P1 coupling misalignment, results of 3D particle-in-cell (PIC) simulations (Appendix C) using WarpX [38] of guiding and acceleration in our 20 cm plasma waveguides are shown in Fig. 2(e)-(g). We consider P1 coupling transverse offsets of 0, 10, and 20 μm. Panel (e) plots the transmitted pulse energy fraction $T(z)$ vs. propagation distance for $a_0 = 0.1, 0.3,$ and $3.0$, illustrating (1) poor transmission for a laser intensity ($a_0 = 0.1$) insufficient to support self-waveguiding, but greatly improved for $a_0 = 0.3$, (2) high transmission even at low $a_0$ for pre-ionized index structures, and (3) significant laser pulse energy depletion into plasma waves, showing transmission consistent with our measured throughput of <40% at high energy. Panel (f) shows that increasingly off-axis P1 coupling leads to increased guided mode centroid oscillation and reduced accelerated charge. Panel (g) shows accelerated bunch spectra. While all are in the range ~1-2 GeV (limited by dephasing, since $L_d < L_{guide}$, and consistent with the measurements of Fig. 2(d)), the accelerated charge (integral of the spectra) decreases significantly with P1 coupling offset, as also seen in Fig. 2(f) and in the measurements (Fig. 2(a)-(d)). Interestingly, while the measured electron spectra (Fig. 2(d)) show quasi-monoenergetic modulations, the simulations show wide, continuous spectra except for the 20 μm coupling offset case. We will discuss the implications of these features in the context of the highest acceleration results, presented next.

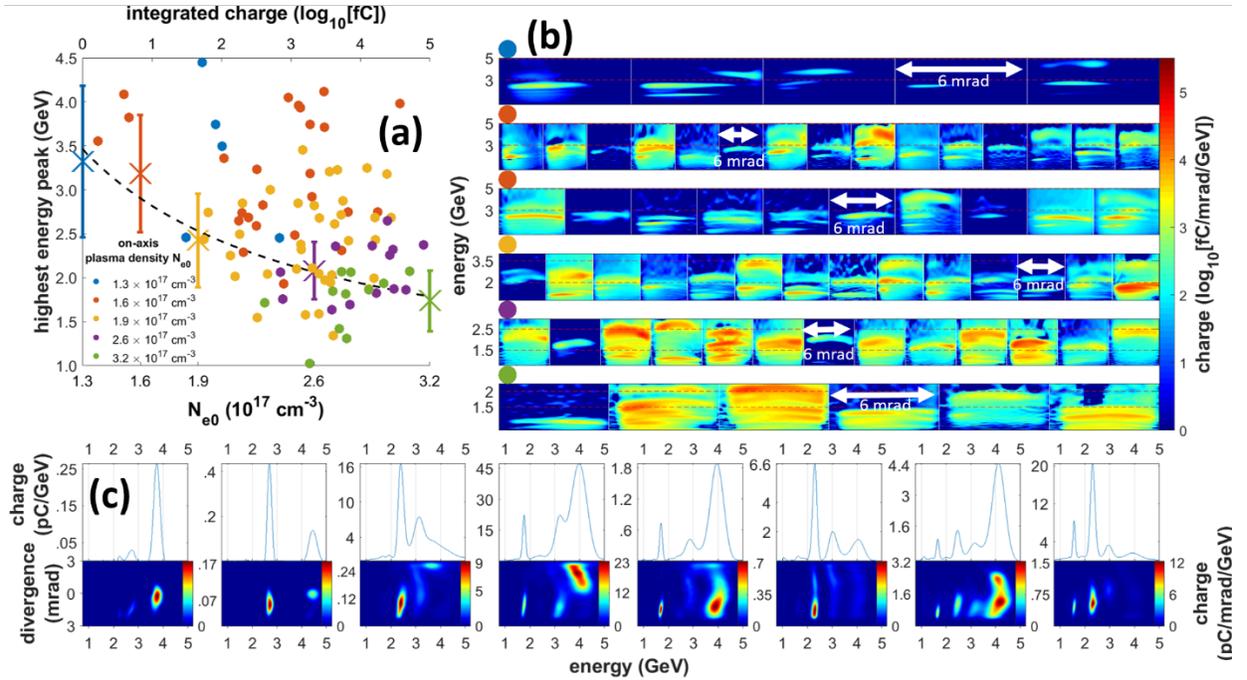

**Figure 3.** Density scan of 20 cm plasma waveguide showing increasing peak bunch energy for decreasing on-axis waveguide density $N_{e0}$. **(a)** Peak bunch energy and associated charge in the highest energy peak for each shot (coloured dots, plotted vs charge) and average peak energy (crosses, plotted vs. $N_{e0}$). The dashed curve is a fit to $\Delta W \propto N_{e0}^{-1}$. **(b)** Angle resolved electron spectra corresponding to coloured dots in (a). **(c)** Spectrum lineouts and angle-resolved spectra for shots with highest energy bunches. The two leftmost panels are for $N_{e0} = 1.3 \times 10^{17}$ cm$^{-3}$ and the rest are for $N_{e0} = 1.6 \times 10^{17}$ cm$^{-3}$.

We achieve electron bunch acceleration up to a maximum of ~5 GeV by operating at lower plasma density. Figure 3 shows results from a plasma waveguide central density scan for $N_{e0} = (1.3 - 3.2) \times 10^{17}$ cm$^{-3}$ and laser energy 11 J, spanning the transition from monomode to low order multimode guiding shown in Fig. 1(g). Notably, using higher laser energy of 15 J under these conditions resulted in reduced peak electron energies, as will be explained below. Figure 3(a) plots peak bunch energy and associated charge (in the highest energy peak) for all shots in the density scan. This is overlaid by average peak energy vs. $N_{e0}$, showing good agreement with the expected $\Delta W \propto N_{e0}^{-1}$ scaling. The charge measurements in all panels represent a lower bound due to the use of the 1 mm entrance slit on the magnetic spectrometer, employed to increase the energy resolution. The actual accelerated charge on a given shot could be up to ~100 times higher for any given shot (see Appendix A). Angle-resolved spectra are shown in Fig. 3(b), while spectrum lineouts and angle-resolved spectra for the highest energy shots (for $N_{e0} < 2 \times 10^{17}$ cm$^{-3}$, where $L_d > L_{guide} = 20$ cm) up to ~ 5 GeV are plotted in Fig. 3(c). The beams have ~milliradian divergence and the narrowest quasi-monoenergetic peaks have relative energy width of ~15%. It is clear that even for fixed nominal waveguide and laser parameters, there are shot-to-shot bunch spectrum and charge variations. As in the higher density experiment of Fig. 2, these variations are partially attributable to fluctuations in P1 pointing. But now, because $L_d > L_{guide}$, the highest energy electron spectra are significantly more sensitive to the axial location of electron injection, which is itself affected by the laser coupling offset and the details of the longitudinal variation of the plasma waveguide.

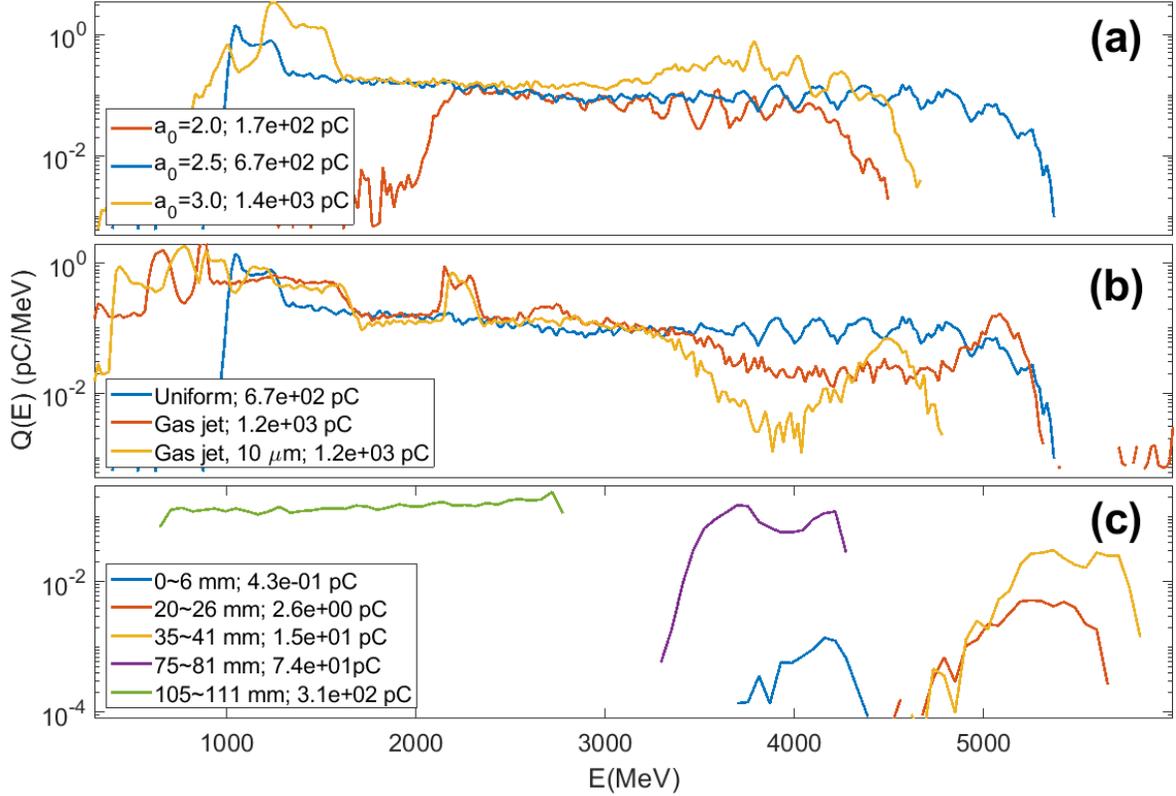

**Figure 4.** Particle-in-cell simulations using WarpX [38] of electron acceleration in 20 cm long plasma waveguide formed in 95% H$_2$ and 5% N$_2$. The guide is initialized with hydrogen fully ionized and nitrogen ionized to N$^{5+}$ (see Appendix C). The charges shown are for electrons with energy > 300 MeV. **(a)** On-axis coupling of $a_0 = 2.0, 2.5,$ and 3.0 pulses into axially uniform waveguide with $N_{e0} = 1.7 \times 10^{17}$ cm$^{-3}$ ($L_d = 27$ cm). **(b)** Coupling of $a_0 = 2.5$ pulse (*i*) on-axis into uniform waveguide with $N_{e0} = 1.7 \times 10^{17}$ cm$^{-3}$ , (*ii*) on-axis into waveguide with on-axis waveguide density $N_{e0}$ proportional to the longitudinal gas jet profile of Fig. 1(c) at 3 mm above the nozzle, and (*iii*) 10 μm off-axis into the same profile as (ii). **(c)** On-axis coupling of $a_0 = 2.5$ pulse into uniform waveguide with $N_{e0} = 1.7 \times 10^{17}$ cm$^{-3}$ with restricted 6 mm sections of 5% N$^{5+}$ placed successively at locations shown in the legend. The short 5% dopant region consists of a 1-mm upramp, a 4-mm plateau and a 1-mm downramp.

Insight into the effects of beam pointing and axial waveguide non-uniformity on electron injection is obtained from the WarpX particle-in-cell simulations ([38], Appendix C) of Fig. 4. Accelerated bunch spectra for an axially uniform waveguide with $N_{e0} = 1.7 \times 10^{17}$ cm$^{-3}$ ($L_d = 27$ cm) are shown in Fig. 4(a) for on-axis coupling of P1 with $a_0 = 2.0 - 3.0$. Broad, relatively flat spectra are seen with highest energy in the range 4-5 GeV. While this agrees with the maximum energy of the experiments, the experimental spectra show peaks with quasi-monoenergetic structure. The broad spectra observed in the simulations originate from continuous ionization injection during guided propagation in the uniform waveguide, while more localized injection evidently occurs in the experiment. The reduction in peak electron energy in going from $a_0 = 2.5$ to $a_0 = 3.0$ in Fig. 4(a) is consistent with our observations of reduced acceleration at higher laser energy. This is due to laser depletion-induced dephasing, as discussed in Appendix C.

Shot-to-shot variation in localized injection occurs from a combination of P1 pointing fluctuations and axial nonuniformity of the plasma waveguide, with the latter originating from axial nonuniformity in the jet's gas density, as measured in Fig. 1(c). As OFI-driven plasma ionization and heating is independent of gas density, the plasma waveguide transverse shape is $z$-invariant but the waveguide on-axis density $N_{e0}$ (and the He-like nitrogen density) varies, and this affects electron injection. This is seen in Fig. 4(b), where the electron spectra are different for a uniform guide and one where the baseline density follows the measured gas jet density profile of Fig. 1(c) (at 3 mm above the nozzle). Off-axis coupling of P1 by 10 μm into the gas jet index structure changes the electron spectrum even more substantially, both in distribution and maximum energy, with the appearance of a quasi-monoenergetic peak centred near 4.5 GeV with ~10% spread, suggesting localized injection in this case.

Detailed inspection of simulation results shows that such localized injection can be triggered by multiple contributing factors during the experiment. First, non-uniformity along the waveguide can lead to ionization injection assisted by sharp density gradients [40]. Second, toward the end of the plasma channel, laser pulse depletion may decrease the peak intensity below the barrier-suppression-ionization threshold of $N^{5+}$ of $a_0$~2.2. Third, transverse offset coupling of the drive laser pulse into the plasma waveguide can have several effects. One is that the transverse oscillation of the laser pulse centroid at the beginning of the waveguide will drive transverse oscillating plasma wakes, suppressing electron injection (see Appendix C). Another is that the beating of the multiple transverse modes excited by off-axis injection can result in intensity spikes that trigger localized ionization injection and minima that suppress it. Consideration of all these effects points to the advantage of localizing the dopant region by design to improve the energy spread and stability of the accelerated electron beam.

When electrons are injected over a restricted longitudinal region, the final electron energy spread can be <10%, with total charge on the order of 10 pC, as illustrated in the simulations of Fig. 4(c). Here, nitrogen dopant is confined to a short 6-mm section successively placed at 5 locations along an axially uniform plasma waveguide. Note that the maximum energy of > ~ 5 GeV occurs for electron injection 2-4 cm after the beginning of the waveguide rather than near the guide entrance. This is because some propagation distance is needed for stabilization of the injected mode and its driven wakefield, and for intensity enhancement by self-steepening, which triggers ionization injection. The short section length is chosen to correspond to the smallest scale of axial gas density variation in Fig. 1(c), qualitatively modeling the effect of axial gas density variations on producing quasi-monoenergetic structure in our measured electron bunches.

## IV. CONCLUSIONS

We have presented results from the first all-optical laser wakefield acceleration experiment to generate multi-GeV electron bunches, here up to 5 GeV. Key to these results are the development of the self-waveguiding technique for generation of meter-scale low density plasma waveguides, as well as long supersonic gas jets that make guiding and acceleration possible.

Optical guiding of relativistically intense ($a_0 > 1$) pulses over such long distances highlights a number of experimental factors whose control will lead to far more stable and monoenergetic accelerated electron bunches: (1) Pointing stability of the drive laser pulse is of paramount importance: if the plasma waveguide supports modes higher than the fundamental, they will be excited by off-axis coupling of the drive pulse, and the resulting guided beam centroid oscillations will excite asymmetric plasma wakes leading to poor or no acceleration. In guides supporting only the fundamental mode, drive laser pointing fluctuations lead to shot-to-shot variation in guided mode intensity and in the driven plasma wakefields. While methods for drive pulse pointing stabilization are ultimately essential, in their absence the entrance section of our plasma waveguides can be designed to perform a mode-filtering function. (2) Fine control of the gas type and axial density distribution is essential to ensure control over the location of electron injection and of the guiding and acceleration properties. This can be implemented by introducing specialized axial sections of the gas jet and by tighter tolerances in the nozzle fabrication to ensure local gas flow uniformity. (3) Over these long guided propagation distances, the evolution of the pulse envelope must be considered in the accelerator design. From initial self-steepening to later red-shifting and stretching owing to depletion—and depletion-induced dephasing—this evolution shows that design of the input drive pulse envelope and peak field $a_0$ can be used to optimize accelerator performance.

## ACKNOWLEDGEMENTS

The authors thank Elaine Taylor and Matthew Le for useful discussions and technical assistance. This research used the open-source particle-in-cell code WarpX (https://github.com/ECP-WarpX/WarpX), primarily funded by the US DOE Exascale Computing Project. We acknowledge all WarpX contributors. High performance computing support to the University of Maryland was provided through the Office of Naval Research (N00014-20-1-2233). Computing resources for preliminary simulations at Oxford were provided by STFC's Scientific Computing Department's SCARF cluster. JJR acknowledges the support of DOD ONR award N00014-20-1-2842. This work was supported by the US Department of Energy (DE-SC0015516, LaserNetUS DE-SC0019076 / FWP#SCW1668, and DE-SC0011375), and the National Science Foundation (PHY1619582 and PHY2010511).

# APPENDIX A: EXPERIMENTAL SETUP

## 1. Measurement of axial density distribution of gas jet

The gas jet system consists of a Mach 4 supersonic slit nozzle fed by five high pressure solenoid valves (backed in the range 200-500 psi), triggered by a driving circuit synchronized to the laser pulses. The nozzle is 20-cm long with 2-mm wide orifice. The valves are held open for ~10 milliseconds to allow the nozzle flow to reach steady state before the arrival of the $J_0$ Bessel beam pulse. The gas feed for the solenoids is pure $H_2$ or a 95%/5% $H_2/N_2$ mix.

The gas jet density profile along the nozzle is measured by imaging the hydrogen recombination fluorescence induced by ionization of the pulsed gas sheet by the $J_0$ beam (Fig. A1(a)). The sharp ends in the image are located at the gas sheet-vacuum boundaries. Gas jet fluorescence images are calibrated by similar images of $J_0$-induced ionization of a known uniform backfill gas density above the nozzle without the valves pulsing (Fig. A1(b)). Images are taken through a 656 nm bandpass filter (bandwidth 10 nm) to restrict recombination imaging to the H-alpha line. Gas jet fluorescence images (Fig. 1(a)) were then interpolated with reference to the uniform density backfill images (Fig. 1(b)). Backfill fluorescence image lineouts for a range of gas pressures are shown in Fig. A1(c). The process is repeated for different vertical displacements between the $J_0$ beam axis and jet orifice to yield the axial density profiles plotted in Fig. 1(c) of the main text.

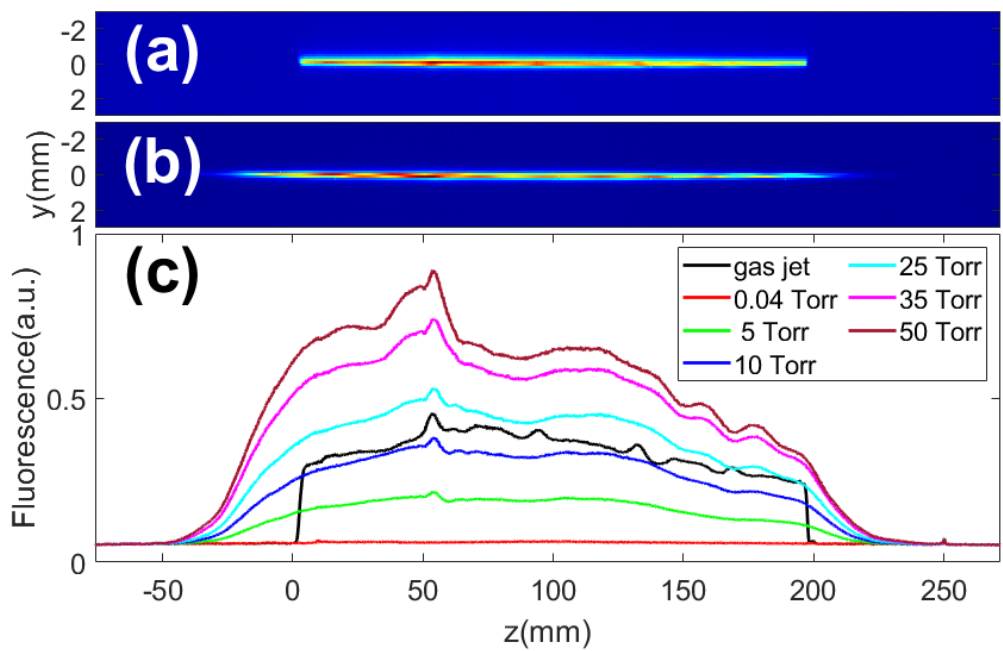

**Figure A1**. Measurement of longitudinal gas density profile. (a) Image of hydrogen plasma fluorescence from the gas jet with 250 psi backing pressure. The $J_0$ beam axis is 3 mm above the nozzle orifice. (b) Image of hydrogen plasma fluorescence in 50 Torr backfill. (c) Lineouts of hydrogen fluorescence for gas jet and hydrogen backfill of various pressures. The bumps in the gas jet fluorescence lineout (black curve) are from gas density variations from nozzle throat width variations and a local intensity bump in the $J_0$ beam focus (near 50 mm).

## 2. Two-color interferometry for measurement of electron density and neutral gas profiles

As discussed in ref. [31], 2-color interferometry (interferometric probe pulses at λ=800nm and λ=400nm as shown in Fig. 1) was required to extract the electron density and neutral gas density profiles in the same shot. These profiles were used for waveguide mode analysis (Appendix B) and for PIC simulations of self-waveguiding and acceleration (Appendix C). Because the low gas and plasma densities imposed very small phase shifts on the probe pulses, noise errors were minimized using many shot averaging in backfill [31]. Sample profiles of plasma and neutral density at 2.5 ns delay after P2 (the index-structuring pulse) is shown in Figure A2. For use in simulations, the extracted Abel-inverted plasma and neutral density profile is fit piecewise with an 8th order polynomial multiplied by an exponential function. To simulate channels formed at different on-axis plasma densities, the profile is normalized to the background gas density ($N_{bg}$) and scaled to the appropriate background level. Based on our extensive 2-color probe measurements of the plasma/neutral expansion [31], the waveguide central density $N_{e0}$ scales with background gas density $N_{bg}$, while the profile shapes do not significantly depend on $N_{bg}$.

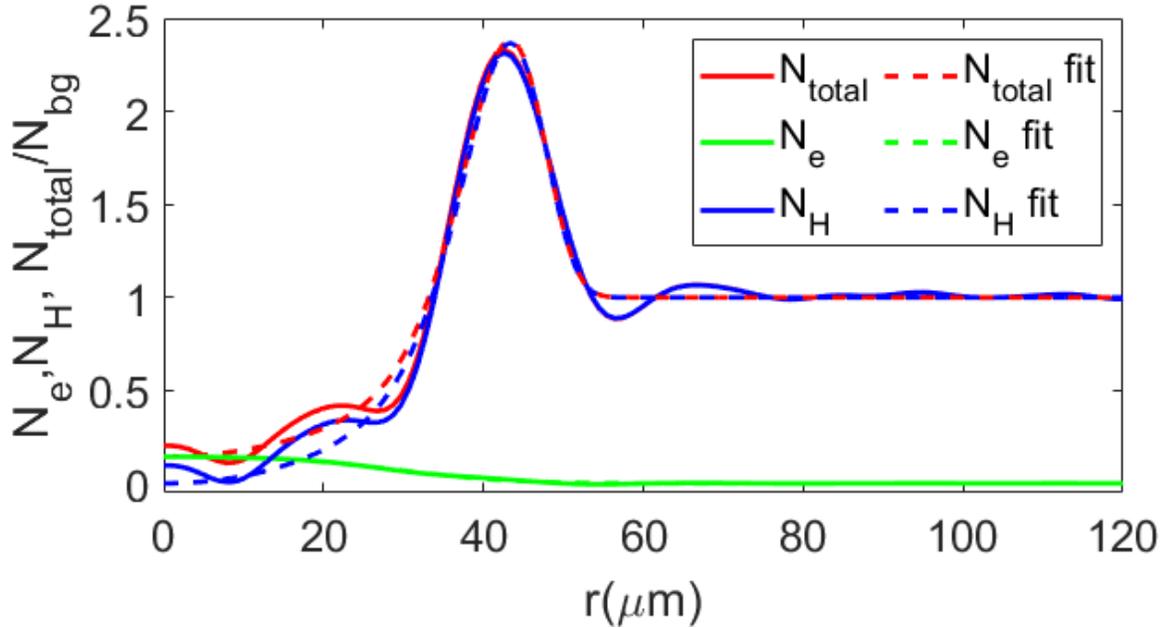

**Figure A2.** Sample measured (solid lines) and fit (dashed lines) profiles of electron density $N_e$, atomic hydrogen density $N_H$, and total density $N_{total} = N_H + N_{H^+}$ at 2.5 ns delay after the index-structuring pulse P2, normalized to the background gas atomic density $N_{bg}$ (here $6.6 \times 10^{18}$ cm$^{-3}$). Note that the concave $N_e$ profile shown here generated by OFI cannot support waveguiding; injection of P1 at 2.5 ns ionizes the neutral density cylindrical shock wall to generate the cladding of the plasma waveguide, as shown in [31] and discussed in the main text.

## 3. Electron Spectra Charge Calibration

Electron beams passed through the magnetic spectrometer's 1 mm (~0.33 mrad) lead entrance slit and were bent in a 0.85 T field of 30 cm extent. Kodak Biomax MS Lanex [41] was used as the scintillating screen LA2 (see Fig. 1), with the fluorescence imaged by a Andor Zyla 4.2 camera. We note that the Lanex response has been observed to decrease approximately linearly with electron energy up to 1.5 GeV [42]. However, we did not have the means to verify the calibration for our Lanex screen at multi-GeV electron energies, so we used the low-energy (40 MeV) calibration from [41]. Assuming the Lanex response trend of [42] leads to undercounting of the signal and a conservative estimate of the bunch charge.

More important to the measurement of bunch charge is our use of the narrow 1 mm, 0.33 mrad slit to aperture the beam. With ~milliradian divergence, the electron beams produced in this experiment overfill the slit. Furthermore, as seen in the electron spectra of Fig. 2 and Fig. 3, there is shot-to-shot variation in the electron beam pointing, resulting in misalignment between the beam and slit, where the signal is seen to be cut off by the edges of the magnets.

While a precise measurement of the charge was not possible with our setup, we improved the lower bound on the charge estimate by using the measured beam divergence to estimate the percentage of the beam blocked by the slit. For a given shot, we assumed the electron beam profile was Gaussian, then used the measured divergence to calculate the beam size. From this, we found the percentage of the charge that would be transmitted into the spectrometer for a perfectly aligned electron bunch. This ratio is used for each shot as an additional calibration factor to estimate the actual charge. This still underestimates the charge, sometimes significantly, for electron bunches not perfectly aligned through the slit.

**APPENDIX B: QUASI-BOUND MODE ANALYSIS**

Due to the finite thickness and height of their cladding, optically generated plasma waveguides are leaky to varying degree, and their modes are quasi-bound [16]. While excessive leakiness can lead to rapid exponential energy attenuation, a guide with a well-bound fundamental mode but leaky higher order modes is desirable to help regularize the laser profile for generation of well-behaved laser wakefields. For a given waveguide transverse profile, the leakiness of a quasi-bound mode is quantified by the $1/e$ energy decay length, $L_{1/e}$ [16]. For application to LWFA, the plasma waveguide should be designed to ensure that the attenuation length of the fundamental mode satisfies $L_{1/e} > L_d, L_{depl}$, the dephasing and depletion lengths in the LWFA process.

The quasi-bound mode structure and attenuation lengths for a radially symmetric plasma profile are found by solving the cylindrical Helmholtz equation for $(p, m)$ modes, where $p = 0, 1, 2, ...$ and $m = 0, 1, 2, ...$ are radial and azimuthal indices,

$$\frac{d^2\mathcal{E}}{ds^2} + \frac{1}{s}\frac{d\mathcal{E}}{ds} + \left(n^2(s) - \frac{\beta^2}{k_0^2} - \frac{m^2}{s^2}\right)\mathcal{E} = 0 \ , \tag{B1}$$

where $E(r, z) = \mathcal{E}(r)e^{i\beta z}$ is the electric field, $k_0$ is the vacuum wavenumber, $\beta$ is the longitudinal propagation number, $s = k_0 r$ is the dimensionless radial coordinate, $n(s)$ is the refractive index profile corresponding to the plasma profile. We identify the quasi-bound modes by solving Eq. (B1) for a range of $\beta' = \beta/k_0$ for fixed azimuthal mode number $m$ and using the $\mathcal{E}(r)$ solutions to identify the maxima of $\eta(\beta') = (|\mathcal{E}_{vacuum}|^2 A)^{-1} \int_A |\mathcal{E}|^2 dA$, where $A$ denotes the waveguide cross-section. For a given $m$, each maximum of $\eta$ identifies the longitudinal wavenumber $\beta$ corresponding to a radial mode $p$. Furthermore, the full width at half maximum, $\Delta\beta$, of the resonance peaks around these maxima gives the attenuation length of each mode: $L_{1/e} = 1/\Delta\beta$ [16]. This is illustrated in Fig. B1(a), were we find the (0,0) and (0,1) modes for the plasma density profile of (a), corresponding to the $N_{e0} = 3.2 \times 10^{17}$ cm$^{-3}$ waveguide of Fig. 1(g)(ii). In Fig. B1(b), the plot of $\eta(\beta')$ shows resonance peaks corresponding to these modes.

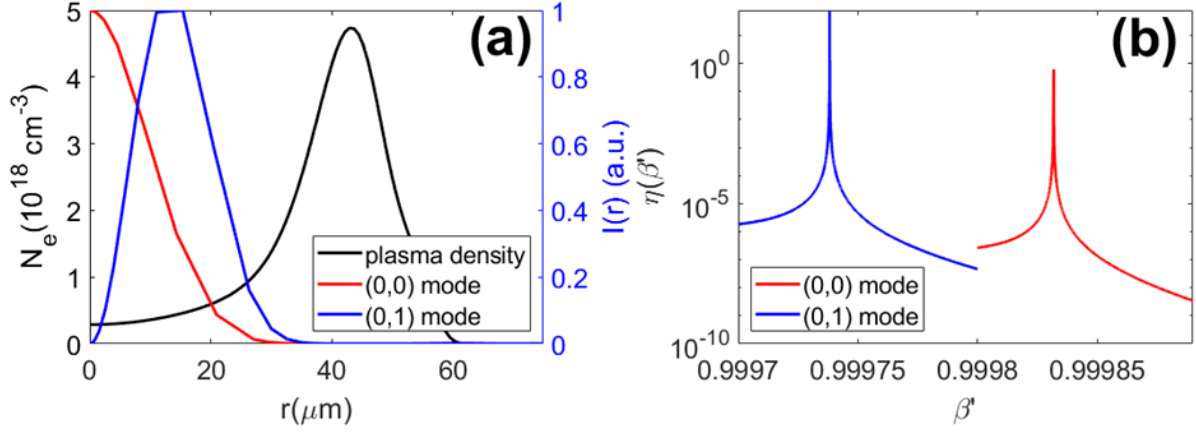

**Figure B1. (a)** Calculated $(p,m) = (0,0)$ and $(0,1)$ modes calculated for waveguide of Fig. 1(g)(ii) with $N_{e0} = 3.2 \times 10^{17}$ cm$^{-3}$. **(b)** Plots of $\eta(\beta')$ vs. $\beta'$ showing the resonances corresponding to the $(0,0)$ and $(0, 1)$ modes.

In general, the fundamental mode will have the longest attenuation length, while the higher-order modes will leak more quickly. However, as illustrated in Fig. 1(g)(ii), guides with high and thick plasma walls may sustain high-order modes for tens of cm. The co-propagation of different modes (with different longitudinal propagation wavenumbers $\beta_j, \beta_k$) leads to mode-beating at a spatial interval $\Lambda = 1/(\beta_j - \beta_k)$, as measured in our paper demonstrating self-waveguiding [31].

**APPENDIX C: PARTICLE-IN-CELL SIMULATIONS**

The particle-in-cell simulations of this paper are performed the with quasi-cylindrical code FBPIC [37] and WarpX [38] in fully 3D geometry. Below we present the simulation parameters and the corresponding plasma/neutral density profiles in detail.

**1. Simulation of the self-waveguiding process**

Figure 1(d) shows the self-waveguiding process simulated using FBPIC in the lab frame, using the parameters shown in table C1. The plasma/neutral density profile is derived from background density scaling of the two-color interferometry measurement shown in Figure A2. Additional self-waveguiding simulations (see Sec. C.2 below) use WarpX.

**Table C1.** FBPIC [37] simulation parameters. Boundary conditions: 'open' in $z$ and 'reflective' in $r$.

| $a_0$ | pol. | $w_0(\mu m)$ | $\tau(fs)$ | window size $N_r \times N_z$ | grid ($\mu m$) ($\Delta z, \Delta r$) | Number of modes $N_m$ | particles per cell $(z, r, \phi)$ |
|---|---|---|---|---|---|---|---|
| 0.3 | $x$ | 30 | 40 | 150×2000 | 0.05, 1.0 | 2 | 1,2,4 |

**2. Simulations of high intensity guiding and laser wakefield acceleration**

3D simulations of high intensity guiding, wakefield generation and acceleration were performed with WarpX in a boosted frame [38]. The simulation parameters are shown in Table C2, with the plasma and neutral density profiles plotted in Fig. C2. Unless otherwise indicated, all simulations assume a fully ionized plasma waveguide with nitrogen atoms ionized to N$^{5+}$. For the high laser energies of our experiments, we found that these results were indistinguishable from self-waveguiding of pulses in the "refractive index structure" consisting of on-axis plasma surrounded by the neutral hydrogen cylindrical shock wall.

| Table C2. WarpX [38] simulation parameters. P1 transverse coupling offsets are 0, 10, and 20 µm. Boundary conditions: perfectly matching layer at $x$, $y$ and $z$ boundaries. | | | | | | | |
|---|---|---|---|---|---|---|---|
| $a_0$ | pol. | $w_0(\mu m)$ | $\tau(fs)$ | window size $N_x \times N_y \times N_z$ | grid ($\mu m$) $(\Delta x, \Delta y, \Delta z)$ | boosted frame $\gamma$ | particles per cell $(x, y, z)$ |
| 2.0, 2.5, 3.0 | x | 30 | 40 | 256×256×4096 | 1, 1, 0.05 | 10 | 1,1,1 |

In Fig. 2(e), two types of plasma/neutral profiles are used to demonstrate the energy loss due to waveguide generation by self-waveguiding and by plasma wave excitation. In the first case, we use a pre-self-waveguiding "index structure" profile for pure hydrogen composed of the plasma and neutral hydrogen density profiles shown in Fig. C1. In the second case, we use a preionized 95% hydrogen and 5% nitrogen gas mix, with the nitrogen atoms ionized to $N^{5+}$. In both cases the density profile is longitudinally uniform over the 20 cm length of the waveguide.

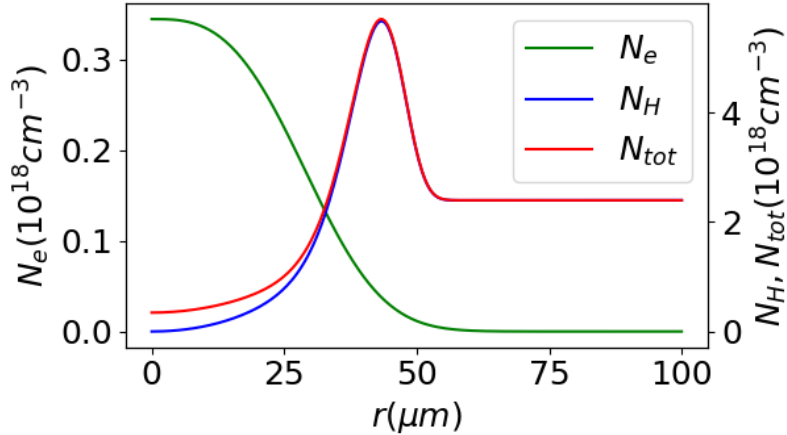

**Figure C1.** Plasma/neutral density profile used in the WarpX simulation of Fig. 2(e).

The effect of longitudinal non-uniformity was modeled by scaling the plasma density profile (Fig. C2(a)) by the measured axial density distribution of Fig. 1(c) (reproduced in Fig. C2(b)).

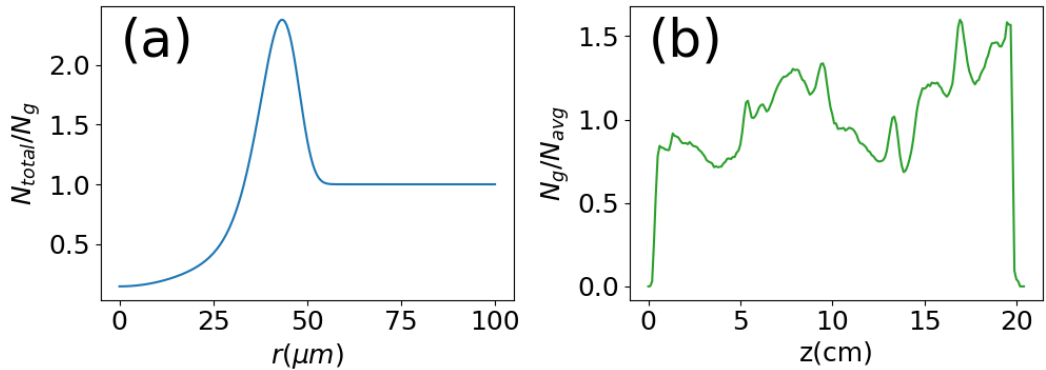

**Figure C2.** (a) Transverse plasma density profile (where $N_{tot}(r) = N_e(r)$ for a fully ionized profile) and (b) measured longitudinal gas density profile of the gas jet (3mm profile from Fig. 1(c)), where $N_{avg}$ is the axial average gas density.

## 3. The effect of pump depletion on electron dephasing

As discussed in prior work, pump depletion can effectively shorten the dephasing length in LWFA [3, 39,43,44]. This is caused by the wake-induced laser red shift that slows down the pulse group velocity and therefore the plasma wake phase velocity. The group velocity of an undepleted laser pulse in a plasma waveguide is [16, 31] $v_{g0}/c \approx 1 - \omega_{p0}^2/2\omega_0^2 - 1/(k_0 w_{ch})^2$, which corresponds to the plasma wake phase velocity $v_{p0}/c$ in the undepleted case. By replacing the initial central laser frequency $\omega_0$ and wavenumber $k_0$ by their instantaneous average in the co-moving simulation frame, $<\omega>$ and $<k>$, one can account for the pump depletion-induced slowdown of plasma wakes in the plasma channel [3, 39,43,44].

We demonstrate this correction by 3D PIC simulation in WarpX and show that pump depletion can reduce the effective dephasing length during propagation in long plasma waveguides. The WarpX simulation parameters are shown in Table C3.

**Table C3.** WarpX parameters for simulation of pump depletion-induced dephasing. Boundary conditions: perfectly matching layer at $x$, $y$ and $z$ boundaries.

| $a_0$ | pol. | $w_0$ ($\mu m$) | $\tau(fs)$ | $N_{eo}$ ($10^{18} cm^{-3}$) | window size $N_x \times N_y \times N_z$ | grid ($\mu m$) $\Delta x, \Delta y, \Delta z$ | Boosted frame $\gamma$ | particles per cell $(x, y, z)$ |
|---|---|---|---|---|---|---|---|---|
| 2.0, 2.5 | $x$ | 30 | 40 | 0.17-0.34 | 256×256×4096 | 1, 1, 0.05 | 10 | 1,1,1 |

Figure C3 shows the mean normalized laser wavenumber $\langle k \rangle/k_0$ in the moving simulation window along the waveguide for several laser and plasma conditions. It is seen that for a given waveguide central density $N_{e0}$, $\langle k \rangle/k_0$ drops increasingly with propagation for larger $a_0$ (increased red shifting), with saturation occurring for the blue curve and incipient saturation for the black and pink curves. Those curves correspond to the higher $N_{e0}$ cases.

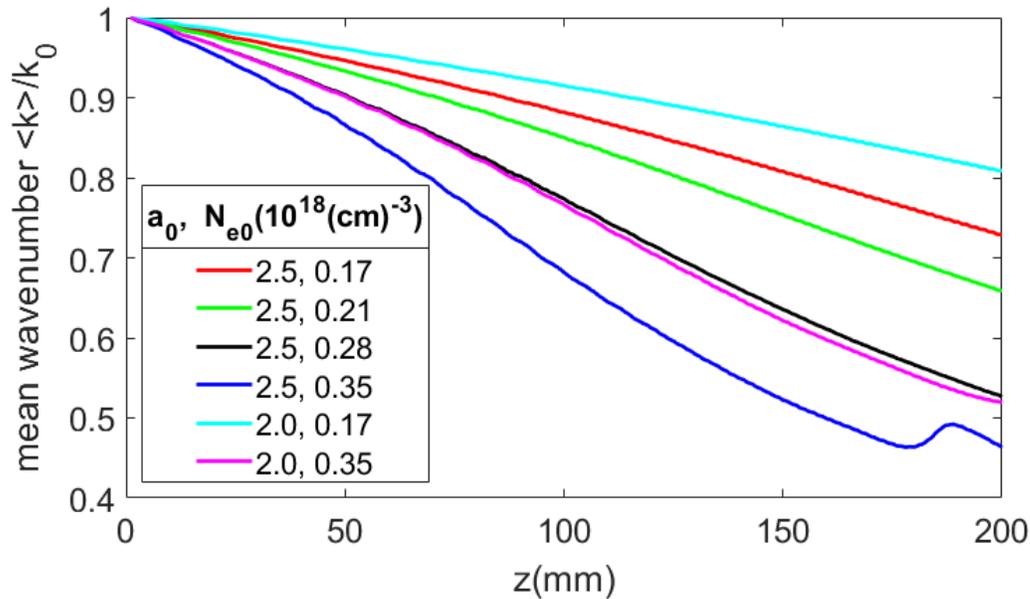

**Figure C3.** Mean normalized wavenumber $\langle k \rangle/k_0$ of the laser pulse along the plasma waveguide.

To see the depletion-induced plasma wake slowdown, in Fig. C4 we plot the wake position (taken as the location $\xi_\phi$ of the peak wake potential after the laser pulse) in the co-moving simulation window with $v_{window} = v_{g0}$. Increasing $a_0$ from 2.0 to 2.5 for fixed waveguide density is seen to increase the wake lag, as does increasing the plasma density for fixed $a_0$. We model this wake lag as a correction $\Delta v_p/c = (v_{p0} - v_p)/c = \omega_{p0}^2/2\omega_0^2 \left(1 - \omega_{p0}^2/<\omega>^2\right) + 1/(k_0 w_{ch})^2 \left(1 - k_0^2/<k>^2\right) + \alpha\sqrt{a_0}(\omega_p/\omega_0)^{3/2}$, with fitting parameter $\alpha = -0.022$. This fit is best for the lower $N_{e0}$ cases (light blue, red, and green curves), but does not capture the fast drop and then saturation for the higher $N_{e0}$ cases (black, pink, and blue curves). The fast drop is caused by the flattening of the local accelerating field by the accelerated electron bunch, and the saturation from pulse lengthening and reduced red shifting accompanying depletion [43,44].

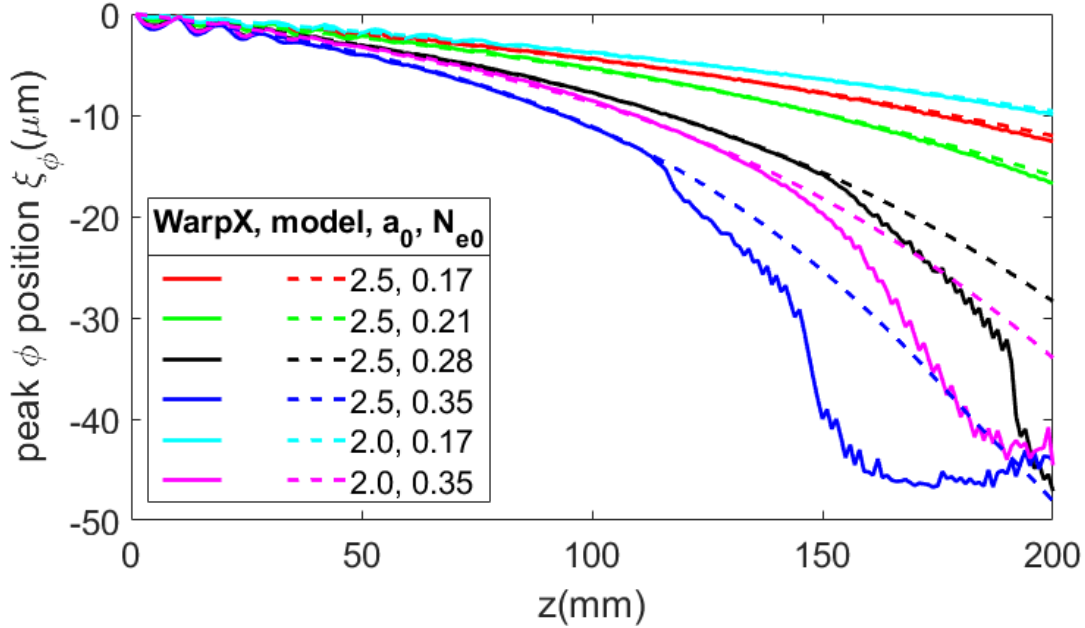

**Figure C4.** Position $\xi$ of the peak potential of the plasma wake bucket following the laser pulse in a frame moving at the group velocity $v_{g0}$ of the undepleted laser pulse. The reduced plasma wave phase velocity effectively shortens the LWFA dephasing length. Solid curves: WarpX simulation. Dashed curves: model fit.

## 4. The effect of drive laser coupling offset on plasma wakefield symmetry

Transverse coupling offset of the drive laser P1 results in centroid oscillations (see Fig. 2(f)) originating from excitation of several modes. This leads to transverse oscillations in the wakefield, as shown in Fig. C5, based on conditions of Fig. 4(b) (gas jet waveguide, $a_0 = 2.5$). Snapshots of the wakefield and its projections simulated using WarpX are shown at $z = 15$ mm and $z = 55$ mm for the cases of on-axis coupling and a transverse coupling offset of 10μm. Here $(x, y, \xi) = (0,0,0)$ corresponds to the laser beam axis and the centroid of the laser pulse in the simulation window moving at the laser pulse group velocity. On-axis coupling leads to transversely symmetric wakefields, with the symmetry persisting over the full length of the waveguide. This is seen most clearly in the projection into the $xy$ plane of the symmetric plasma wake density near the back of the first potential bucket at $\xi = 53$ μm, the centre of the accelerated electron bunch (Fig. C5(a) and (c)). A coupling offset of 10 μm, however, leads to transverse wake asymmetry that persists as long as higher order mode(s) are confined (here, mainly the (0,1) mode), as seen in the same $xy$ projection (for $\xi = 53$ µm) at $z = 55$ mm (Fig. C5(b) and (d)). Also notable is the presence of an on-axis electron bunch of much larger charge in the case of zero P1 coupling offset (compare panels (c) and (d), where in (d) the lower charge bunch is located off-axis). These cases correspond to the two "gas jet" electron spectra shown in Fig. 4(b). This is consistent with the electron bunch fluctuations associated with P1 pointing variations discussed throughout this paper.

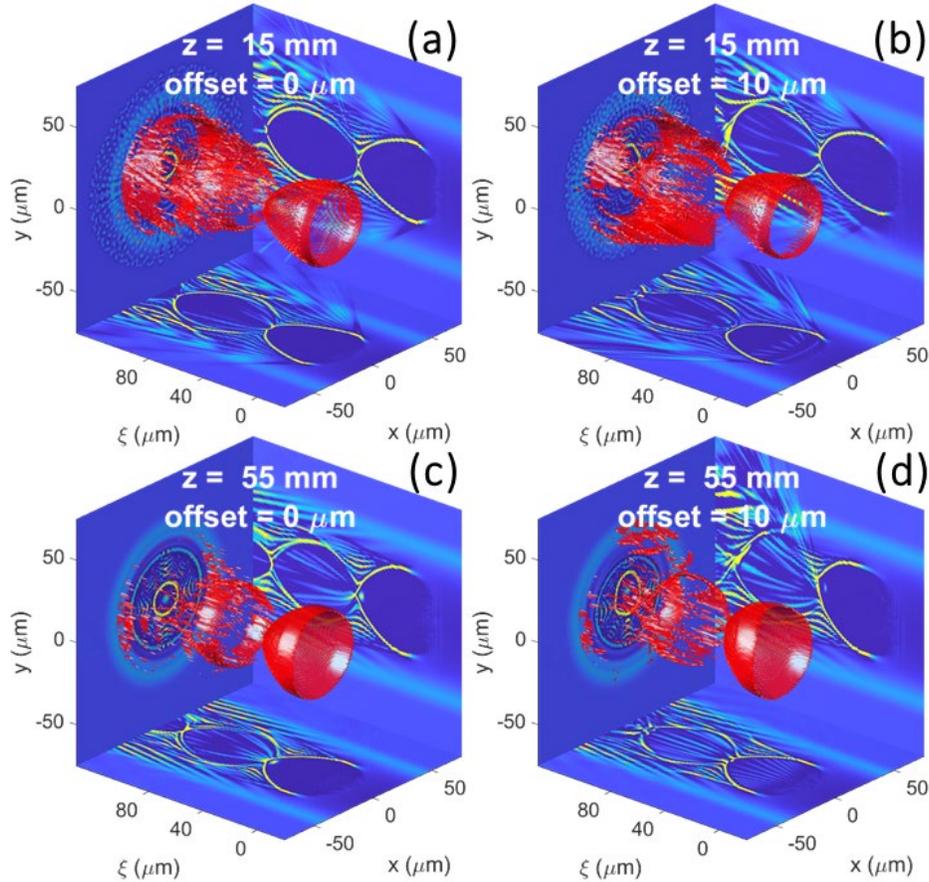

**Figure C5.** WarpX simulation assessing the effect on the plasma wake of off-axis drive pulse (P1) coupling, using the parameters of Fig. 4(b). Here $(x, y, \xi) = (0,0,0)$ corresponds to the laser beam axis and the centroid of the laser pulse. The red contours are for $N_e = 9.4 \times 10^{18}$ cm$^{-3}$, the $xy$ projection is for $\xi = 53$ µm (centre of the accelerated electron bunch), and the $x\xi$ and $y\xi$ projections are for $y = 0$ and $x = 0$ respectively. **(a)** No P1 coupling offset, $z = 15$ mm, **(b)** 10 µm P1 offset, $z = 15$ mm, **(c)** No P1 coupling offset, $z = 55$ mm, **(d)** 10 µm P1 offset, $z = 55$ mm.

*Acceleration to 8 GeV in a Laser-Heated Capillary Discharge Waveguide*, Phys. Rev. Lett. **122**, 084801 (2019).
5. H.- T. Kim, K.-H. Pae, H.- J. Cha, I.- J. Kim, T.- J. Yu, J.- H. Sung, S.- K. Lee, T.- M. Jeong, and J.-M. Lee, *Enhancement of Electron Energy to the Multi-GeV Regime by a Dual-Stage Laser-Wakefield Accelerator Pumped by Petawatt Laser Pulses*, Phys. Rev. Lett. **111**, 165002 (2013).
6. X. Wang, R. Zgadzaj, N. Fazel, Z. Li, S. A. Yi, X. Zhang, W. Henderson, Y. Y. Chang, R. Korzekwa, *et al*., *Quasi-monoenergetic laser-plasma acceleration of electrons to 2 GeV*, Nat. Commun. **4**, 1988 (2013).
7. S. Steinke, J. van Tilborg, C. Benedetti *et al*., *Multistage coupling of independent laser-plasma accelerators*, Nature **530**, 190 (2016).
8. C. B. Schroeder, E. Esarey, C. G. R. Geddes, C. Benedetti, and W. P. Leemans, *Physics considerations for laser-plasma linear colliders*, Phys. Rev. Special Topics Accel. Beams **13**, 101301 (2010).
9. J. Osterhoff, A. Popp, Zs. Major *et al*., *Generation of Stable, Low-Divergence Electron Beams by Laser-Wakefield Acceleration in a Steady-State-Flow Gas Cell*, Phys. Rev. Lett. **101**, 085002 (2008).
10. C. Joshi, *Laser-Driven Plasma Accelerators Operating in the Self-Guided, Blowout Regime*, IEEE Trans. Plasma Sci. **45**, 3134 (2017).
11. C. G. Durfee and H. M. Milchberg, *Light pipe for high intensity laser pulses*, Phys. Rev. Lett. **71**, 2409 (1993).
12. Y. Ehrlich, C. Cohen, A. Zigler, J. Krall, P. Sprangle, and E. Esarey, *Guiding of High Intensity Laser Pulses in Straight and Curved Plasma Channel Experiments*, Phys. Rev. Lett. **77**, 4186 (1996).
13. A. Butler, D. J. Spence, and S. M. Hooker, *Guiding of High-Intensity Laser Pulses with a Hydrogen-Filled Capillary Discharge Waveguide*, Phys. Rev. Lett. **89**, 185003 (2002).
14. W. Lu, M. Tzoufras, C. Joshi, F. S. Tsung, W. B. Mori, J. Vieira, R. A. Fonseca, and L. O. Silva, *Generating multi-GeV electron bunches using single stage laser wakefield acceleration in a 3D nonlinear regime*, Phys. Rev. ST Accel. Beams **10**, 61301 (2007).
15. W. P. Leemans, A. J. Gonsalves, H.-S. Mao, K. Nakamura, C. Benedetti, C. B. Schroeder, Cs. Tóth, J. Daniels, D. E. Mittelberger, S. S. Bulanov, J.-L. Vay, C. G. R. Geddes, and E. Esarey, *Multi-GeV Electron Beams from Capillary-Discharge-Guided Subpetawatt Laser Pulses in the Self-Trapping Regime*, Phys. Rev. Lett. **113**, 245002 (2014).
16. T. R. Clark and H. M. Milchberg, *Optical mode structure of the plasma waveguide*, Phys. Rev. E **61**, 1954 (2000).
17. P. Sprangle, B. Hafizi, J. R. Peñano, R. F. Hubbard, A. Ting, C. I. Moore, D. F. Gordon, A. Zigler, D. Kaganovich, and T. M. Antonsen, *Wakefield generation and GeV acceleration in tapered plasma channels*, Phys. Rev. E **63**, 056405 (2001).
18. C Benedetti, CB Schroeder, E Esarey, WP Leemans, *Quasi-matched propagation of ultra-short, intense laser pulses in plasma channels*, Phys. Plasmas **19**, 053101 (2012).
19. B. D. Layer, A. York, T. M. Antonsen, S. Varma, Y.-H. Chen, Y. Leng, and H. M. Milchberg, *Ultrahigh-Intensity Optical Slow-Wave Structure*, Phys. Rev. Lett. **99**, 035001 (2007).
18